\documentclass{article}
\usepackage{hiph-art}
\usepackage{rotating}

\def\Journal#1#2#3#4{{#1}{\bf{#2}} (#4) #3. }
\def\NPA{{\em Nucl. Phys.} \bf{A}}
\def\PRL{\em Phys. Rev. Lett. }
\def\PLB{{\em Phys. Lett.} \bf{B}}
\def\EPJA{{\em Eur. Phys. J.} \bf{A}}
\def\IEEE{\em IEEE Trans. on Nucl. Science }
\def\NIMA{{\em Nucl. Instr. Meth.} \bf{A}}
\def\ANNP{\em Ann. Phys. }
\def\PPNP{\em Prog. in Part. and Nucl. Phys. }
\def\PRC{{\em Phys. Rev.} \bf{C}}

\volnumber{19} \issuenumber{1} \edyear{2004}                             
\frompage{000} \topage{000}                                              
\recrevdate{1 January 2004}                                              
\def\rightmark{Nucleon resonances in the medium}
\def\leftmark{B. Krusche}
 
\title{In-medium properties of nucleon resonances} 
\authors{ 
{B. Krusche$^{1}$  %
\index{Krusche, B.} 
}\\[2.812mm]
{\normalsize
\hspace*{-8pt}$^1$ Department of Physics and Astronomy, University of Basel,\\ 
4056 Basel, Switzerland\\[0.2ex] 
}}
 
\abstract{Recent experimental results for the in-medium properties of
nucleon resonances are discussed. The experiments were done with the TAPS 
detector at the tagged photon beam of the MAMI accelerator in Mainz. 
Measured was the photoproduction of mesons (final states $\pi^o X$, 
$\eta X$, $2\pi^oX$ and $\pi^o\pi^{\pm}X$) from the nuclei $^{12}$C, $^{40}$Ca, 
$^{93}$Nb, and $^{208}$Pb up to the second resonance region. The results 
were analyzed in view of the in-medium properties of the P$_{33}$(1232), 
the D$_{13}$(1520), and the S$_{11}$(1535) resonances.}

 
\makeindex
\begin{document}
 
\maketitle

\section{Introduction}\label{intro}
QCD at high energies or short scales ($r < 0.1$ fm) is a perturbative theory
with point-like quarks and gluons. However, at larger distances 
the perturbative picture breaks down. In the intermediate range (0.1 fm $< r <$
1 fm) the physics is governed by the excitation of nucleon resonances. This
means that the full complexity of the nucleon as a many body system with 
valence quarks, sea quarks and gluons contributes.
At even larger distances beyond 1 fm, QCD becomes the theory
of nucleons and mesons (pions) and can be treated in the framework of chiral
perturbation theory. Chiral symmetry is at the very heart of low energy QCD.
However, it is well known, that chiral symmetry is
spontaneously broken. This is connected to a non-zero
expectation value of scalar $q\bar{q}$ pairs in the vacuum, the 
chiral condensate. A consequence of the chiral symmetry breaking in the
hadron spectra is the non-degeneracy of parity doublets. 

Model calculations indicate a significant temperature and density dependence 
of the chiral condensate (see e.g. \cite{Lutz_92}). It's melting
is connected with the prediction of a partial restoration of chiral symmetry 
at high temperatures and/or large densities. 
One consequence of the partial chiral symmetry restoration is a density 
dependence of hadron masses. An early prediction for this effect is the 
so-called Brown-Rho scaling \cite{BR_91}.
Evidence for such effects has been searched for in many experiments. An 
example is the search for the predicted shift and broadening of the 
$\rho$-meson in the di-lepton spectra of heavy ion reactions with CERES at 
CERN \cite{Agakichiev_95,Adamova_03}.

\begin{figure}[ht]
\epsfysize=1.5cm \epsffile{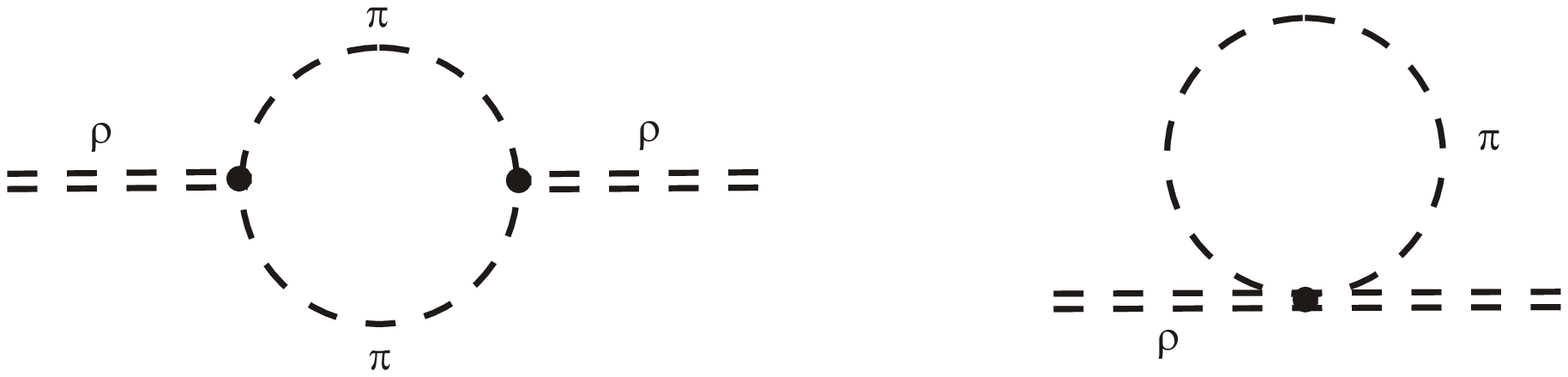}
\hspace*{0.5cm}\epsfysize=1.5cm \epsffile{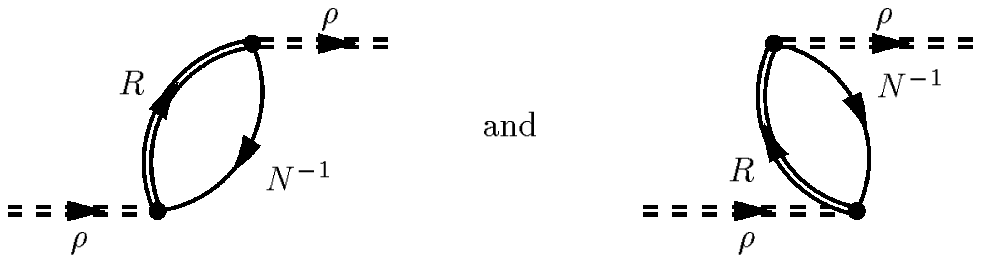}\\
\hspace*{0.5cm}\epsfysize=1.7cm \epsffile{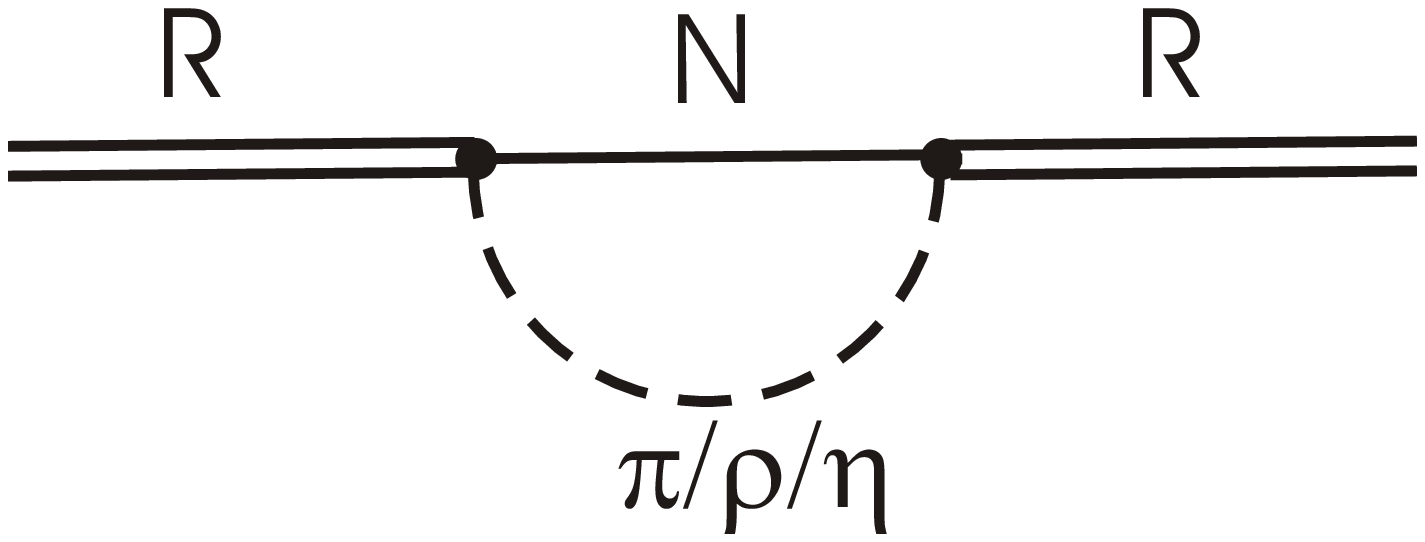}
\hspace*{3cm}\epsfysize=1.7cm \epsffile{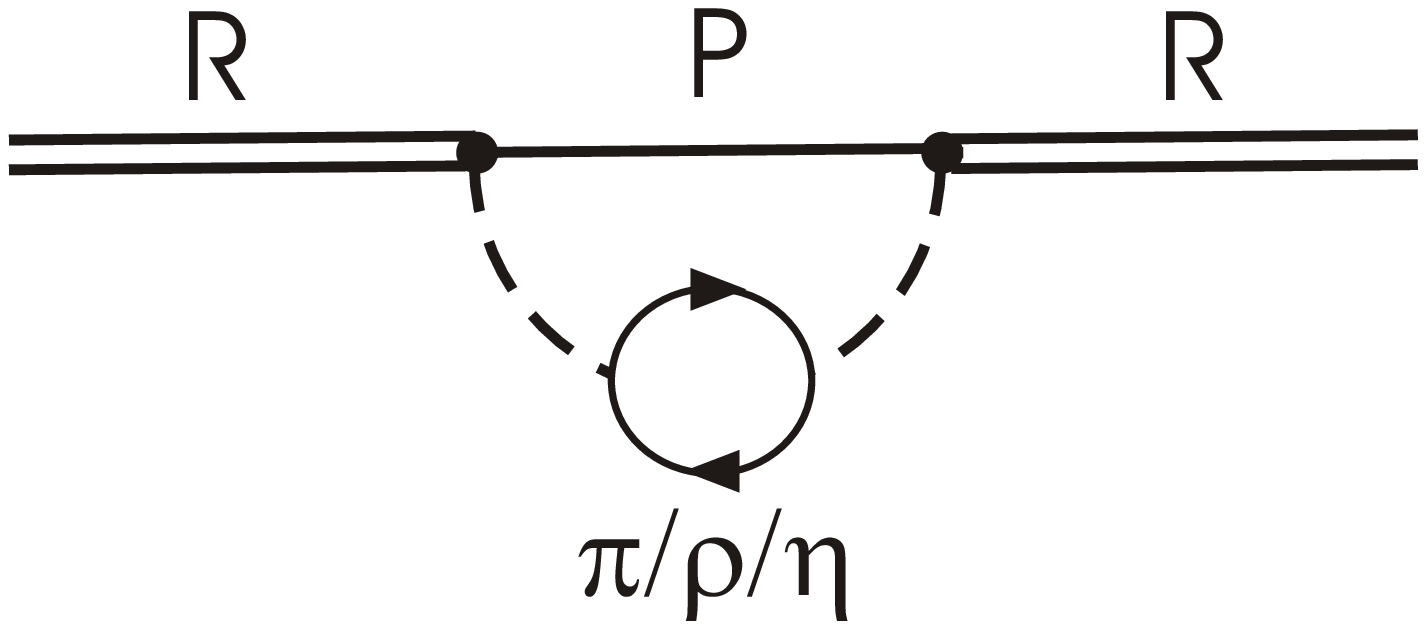}
\vspace*{-0.3cm}
\caption{Self energies from coupling of mesons and nucleon resonances, left
hand side: vacuum, right hand side: in the nuclear medium \cite{Post_04}}
\label{fig:1}
\end{figure}

In-medium modifications of mesons will also influence the in-medium properties 
of nucleon resonances.
Recently, Post, Leupold and Mosel \cite{Post_04} have calculated in a 
self consistent way the spectral functions of mesons and baryons in nuclear 
matter. The most relevant contributions to the 
self-energies are shown in fig. \ref{fig:1}. In the vacuum, mesons couple 
only to meson loops (involving e.g. the pion) and nucleon 
resonances couple to nucleon - meson loops. However, in the medium mesons 
couple also to resonance - hole states. This influences not
only the spectral functions of the mesons, but also the resonances, which in 
turn couple to the modified mesons loops. The predicted effects are in
particular large for the $\rho$ meson and the D$_{13}$(1520) resonance.
The close-by S$_{11}$(1535) resonance is much less effected
\cite{Post_04}.  

During the last few years, the TAPS collaboration has 
studied the in-medium properties of mesons and nucleon resonances and the 
meson - nucleus interactions with photon induced reactions. This program 
covers four different major topics: 
\begin{itemize}
\item{The investigation of resonance contributions to $\eta$, $\pi$, $2\pi$
meson production reactions from light and heavy nuclei 
\cite{Krusche_95b,Roebig_96,Krusche_99,Hejny_99,Kleber_00,Weiss_01,Weiss_03,Krusche_04}.}
\item{The search for $\eta$ - nucleus bound states
($\eta$-mesic nuclei), which would be the ideal testing ground for the
investigation of the $\eta$ - nucleus interaction \cite{Hejny_02,Pfeiffer_04}.}
\item{The investigation of the $\pi\pi$ invariant mass distributions
for $2\pi$ production from nuclei, aiming at the in-medium behavior of the 
'$\sigma$'-meson \cite{Messchendorp_02}.}
\item{The measurement of the resonance shape of the $\omega$ meson in nuclear
matter from its $\pi^o\gamma$ decay.}
\end{itemize} 

Only the first topic will be discussed in this contribution, the status of
the double pion production experiments is reviewed by S. Schadmand and the 
latest results from $\omega$-photoproduction are discussed by D. Trnka
in the same topical issue of Acta Phys. Hungarica A.  

\section{Experiments}\label{techno} 
The experiments were carried out at the Glasgow tagged photon facility 
installed at the Mainz microton MAMI. They used Bremsstrahlung photons 
produced with the 850 MeV electron beam in a radiator foil. 
The meson from the production targets were detected with the electromagnetic
calorimeter TAPS \cite{Novotny_91,Gabler_94}, which consists of more than 
500 hexagonally shaped BaF$_2$ scintillators of 25 cm length corresponding 
to 12 radiation lengths. The separation of photons from massive particles 
makes use of the plastic veto detectors (only charged particles), a 
time-of-flight measurement, and 
the excellent pulse shape discrimination capabilities of BaF$_2$-scintillators.
The identification of neutral mesons ($\pi^o$ and $\eta$) 
is done by a standard invariant mass analysis. Charged mesons and 
nucleons are identified in addition with time-of-flight versus energy 
analyses. Details are summarized in \cite{Krusche_99,Krusche_04}.

\section{Results}\label{results}
Data have been taken for $^2$H, $^{3,4}$He, $^{12}$C, $^{40}$Ca, $^{93}$Nb and 
$^{208}$Pb targets. The data from the deuteron were used as a reference point 
for the elementary cross sections from the quasifree nucleon. 

\subsection{\it The $\Delta$(1232) resonance}
A detailed understanding of the in-medium properties of this
state is necessary for any interpretation of pion photoproduction reactions 
from nuclei. It dominates single pion production in the low energy region up 
to 500 MeV incident photon energies, but it also contributes at higher 
energies via multiple pion production processes and through re-absorption of 
pions. An in-medium broadening at normal nuclear matter density
of roughly 100 MeV has been previously extracted from pion nucleus scattering 
experiments (see e.g. \cite{Hirata_79}). 

\begin{figure}[th]
\centerline{
\epsfysize=4.1cm \epsffile{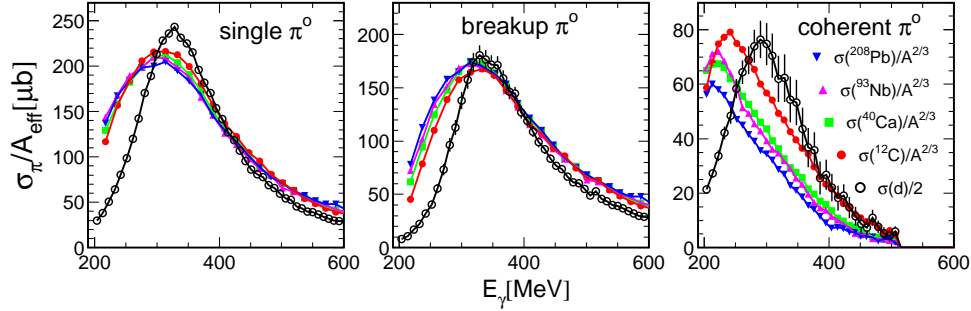}}
\caption{Total cross sections for $\pi^o$ production in the $\Delta$ resonance
from the deuteron and from heavy nuclei. Right hand side: coherent part,
middle: breakup part, left hand side: sum of both 
\cite{Krusche_99,Krusche_02,Krusche_04}. The legend is valid for all three 
pictures.}  
\label{fig:2}
\vspace*{-0.5cm}
\end{figure}

In photon induced reactions on the free proton, single $\pi^o$ photoproduction 
is best suited to study this state \cite{Krusche_03} since the background 
from non-resonant contributions like pion-pole and Kroll-Rudermann terms 
is suppressed for neutral pions. However, for nuclei a further 
complication arises. Neutral pions can be produced in (quasifree) breakup 
reactions, where in simplest plane wave approximation, 
the pion is produced from a single nucleon, which in the process is knocked 
out of the nucleus. However, as long as the momentum transfer is small, 
this process competes with coherent $\pi^o$ production where the amplitudes 
from all nucleons add coherently, the momentum transfer is taken by the entire 
nucleus, and no nucleons are removed. 
The two reaction mechanisms can be separated via their different kinematics.
The total cross sections for the deuteron and the heavy nuclei are summarized 
for both reaction mechanisms \cite{Krusche_99,Krusche_02,Krusche_04}
in fig. \ref{fig:2}. 
Their behavior is quite different. 
The coherent reaction can be approximated for spin $J=0$ nuclei in plane wave 
by:
\begin{equation}
\frac{d\sigma_A}{d\Omega} \propto 
\frac{d\sigma_N}{d\Omega} A^2 F^2(q) sin^2(\Theta^{\star})
\end{equation}
where $d\sigma_A$ is the nuclear cross section, $d\sigma_N$ the elementary
cross section on the free nucleon, $A$ the atomic mass number, $F^2(q)$ the
nuclear form factor depending on the momentum transfer $q$, and $\Theta^{\star}$
the cm polar angle of the pion (for details see \cite{Krusche_02}).  
The observed shift of the peak cross section to low photon energies for heavy 
nuclei is not related to in-medium effects of the $\Delta$ but
is a simple consequence of the interplay between the $F^2(q)$ and 
$sin^2(\Theta^{\star})$ factors. 

\begin{figure}[ht]
\centerline{
\epsfysize=5.cm \epsffile{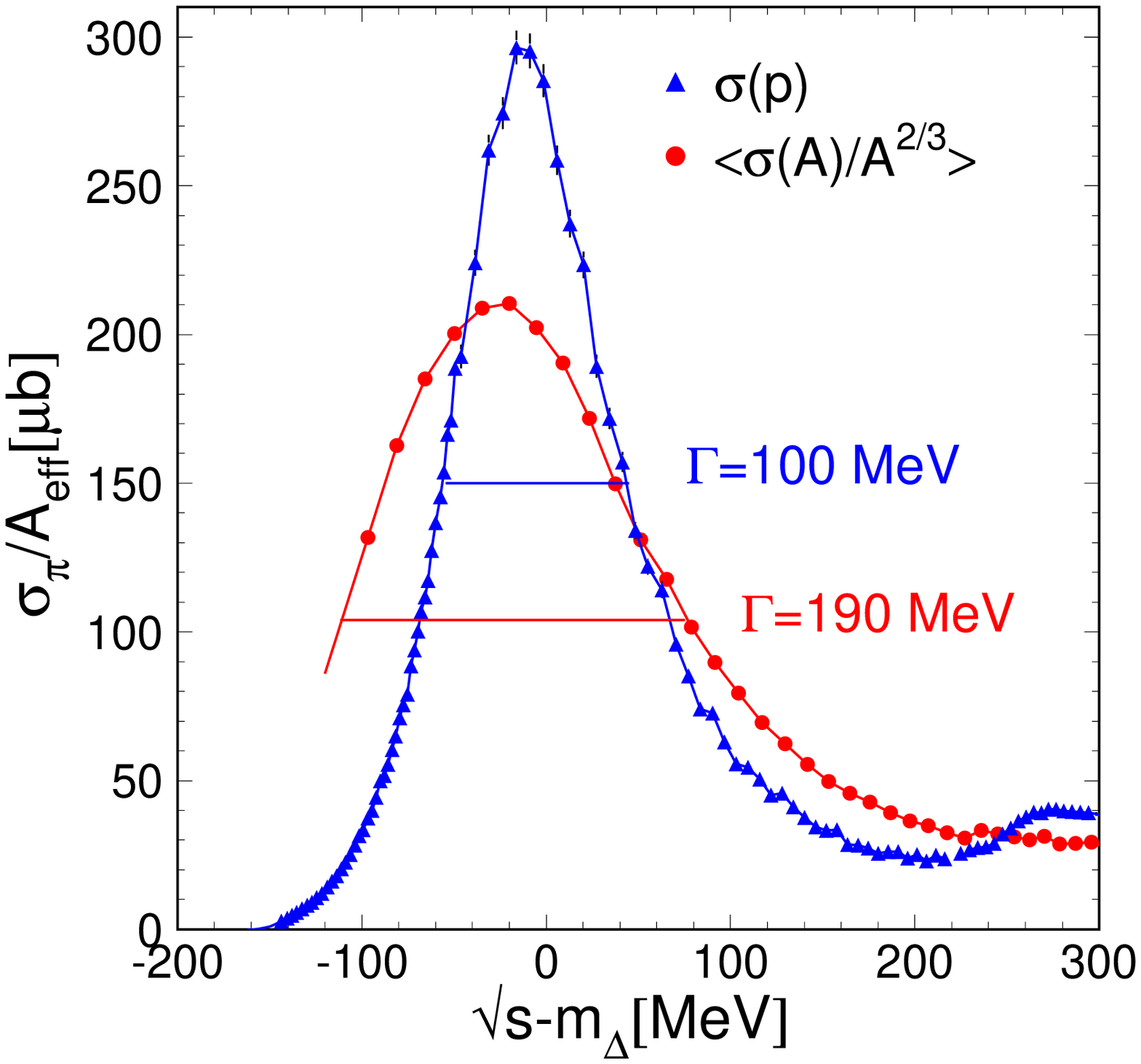}
\hspace*{0.cm}\epsfysize=5.1cm \epsffile{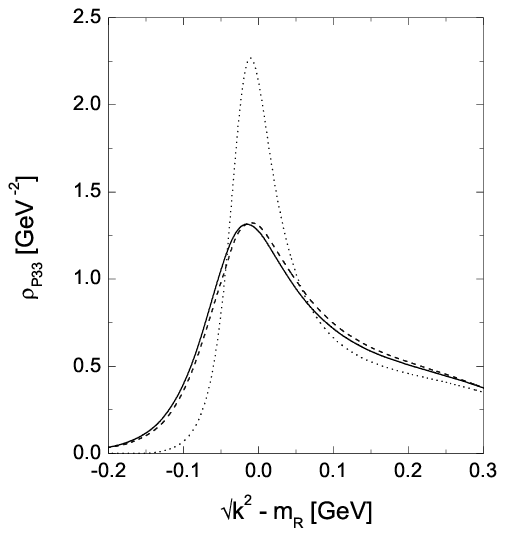}}
\caption{Left hand side: total $\pi^o$ photoproduction from the proton compared
to the average for heavy nuclei. Right hand side: spectral
functions for the $\Delta$ resonance in vacuum and in the nuclear medium
\cite{Post_04}.}
\label{fig:3}
\vspace*{-0.5cm}
\end{figure}

It is tempting to argue, that the quasifree breakup process is best suited to
study the $\Delta$ in the medium. However, quasifree and coherent 
contributions are not independent. They are connected via final state 
interaction (FSI), which was discussed in detail in 
\cite{Krusche_99,Siodlaczek_01}. Siodlaczek et al. \cite{Siodlaczek_01} have 
even argued that for the deuteron the effect of FSI in the breakup process is 
counterbalanced by the coherent process so that the sum of the cross 
sections for the coherent and the breakup part with FSI equals the cross 
section of the pure quasifree process without FSI. A similar effect is 
visible in the angular distributions of single pion photoproduction from 
heavy nuclei \cite{Krusche_04}. The breakup cross section is depleted at 
forward angles where the coherent cross section peaks. 
The cross section for inclusive single $\pi^o$ photoproduction, i.e. the sum
of breakup and quasifree parts can thus serve as a first approximation. It
scales with $A^{2/3}$, which indicates strong
FSI effects. The average over the heavy nuclei is compared in fig. \ref{fig:3}
to the cross section from the free proton. 
The $\Delta$-resonance peak for the nuclei is significantly broadened with
respect to the free nucleon from 100 MeV to 190 MeV. This is in nice agreement 
with the prediction for the in-medium spectral function of the $\Delta$ 
\cite{Post_04}(see fig. \ref{fig:3}, right hand side), which corresponds to
exactly the same broadening.

\begin{figure}[ht]
\centerline{
\epsfysize=4.cm \epsffile{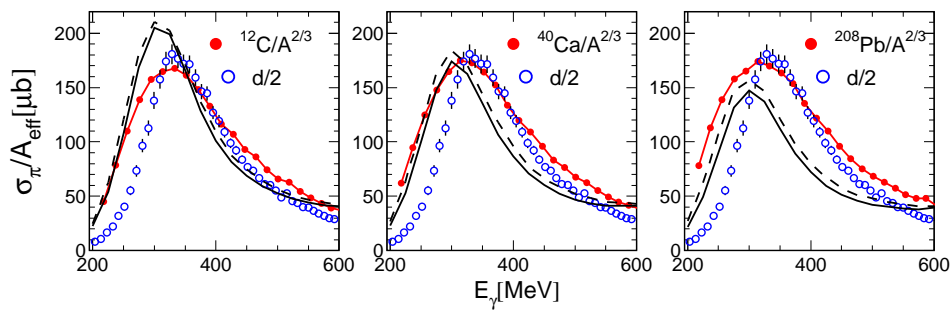}}
\caption{Quasifree photoproduction of single $\pi^o$ mesons from $^2$H
and from heavy nuclei compared to BUU-model calculations 
\cite{Lehr_00,Krusche_04}.}  
\label{fig:4}
\end{figure}

The breakup process is mostly treated in the framework of
nuclear cascade models or transport models. The data are compared in fig.
\ref{fig:4} to calculations in the framework of the Boltzmann-Uehling-Uhlenbeck
transport model \cite{Lehr_00,Krusche_04}. The model includes 
an additional in-medium width of the $\Delta$ of roughly 80 MeV at normal
nuclear matter density.
The calculations reproduce the shift of the rising slope of the $\Delta$ to
lower incident photon energies, but underestimate the falling slope and show a
somewhat different mass number dependence of the peak cross section. 

The results from coherent $\pi^o$ photoproduction from nuclei have been 
analyzed in \cite{Rambo_00,Krusche_02} in the framework of the DWIA 
calculations of Drechsel et al. \cite{Drechsel_99a} which include a 
phenomenological $\Delta$ self-energy. 
The main finding was, that the model with the self-energy fitted to $^4$He
reproduces the data for carbon and calcium so that no
significant mass dependence of the self-energy was found. The self-energy 
itself corresponds to an increase of the width at resonance position
($E_{\gamma}\approx 330$ MeV) of roughly 110 MeV in agreement with the results 
discussed above. The resonance position is slightly upward shifted (by 20 MeV).
This is no contradiction to the excitation functions in figs. 
(\ref{fig:3},\ref{fig:4}). 
The width increase is energy dependent 
(only $\approx$40 MeV at $E_{\gamma}\approx 250$ MeV) so that the net effect in
the excitation functions in figs. (\ref{fig:3},\ref{fig:4}) is a small 
downward shift of the peak position.

\subsection{\it The second resonance region}
\begin{figure}[ht]
\hspace*{0.5cm}\begin{minipage}{7cm}
\begin{turn}{-90.}
\epsfysize=7cm \epsffile{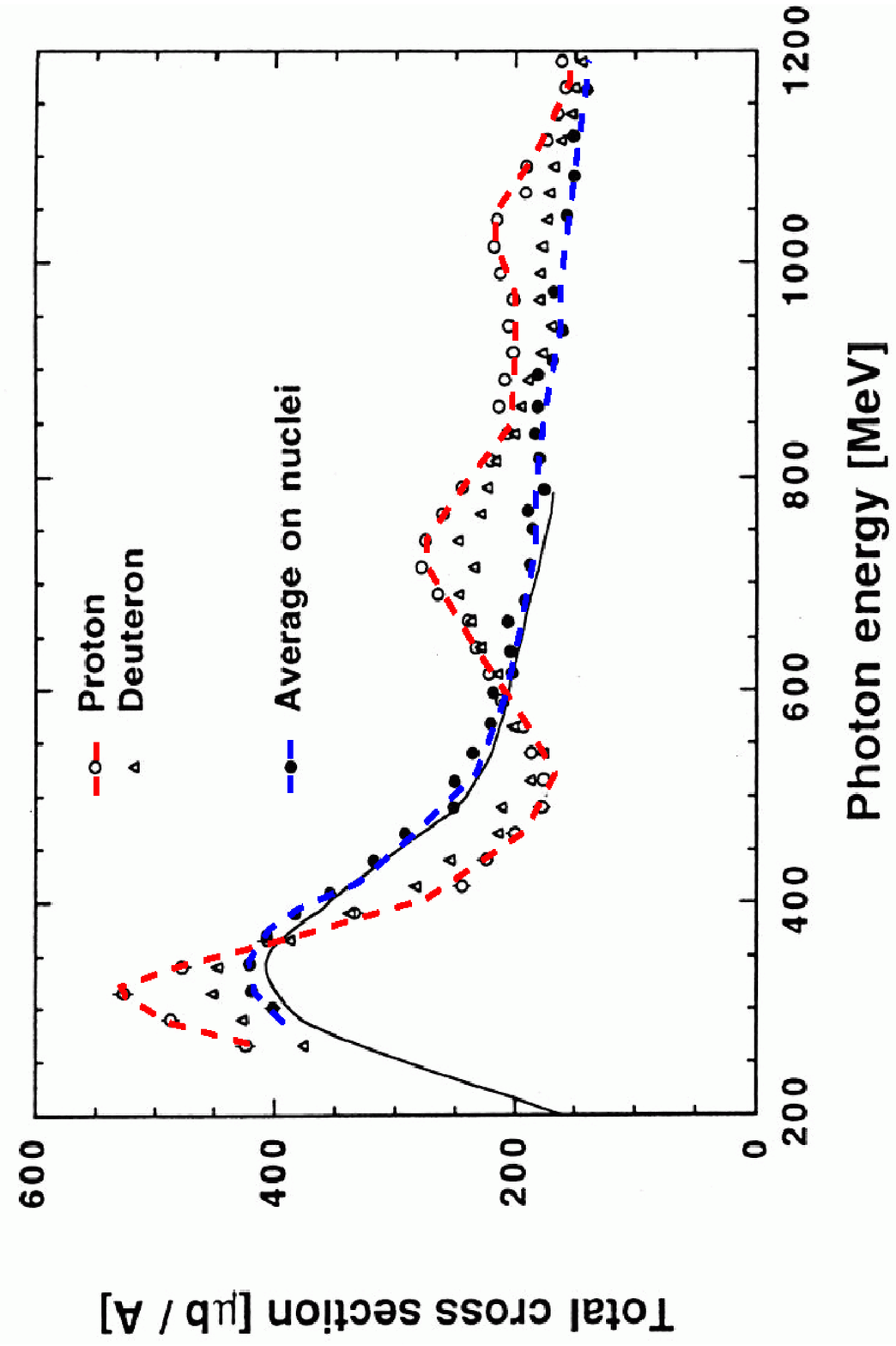}
\end{turn}
\end{minipage}
\begin{minipage}{7.5cm}
\hspace*{0.cm}\epsfysize=4.5cm \epsffile{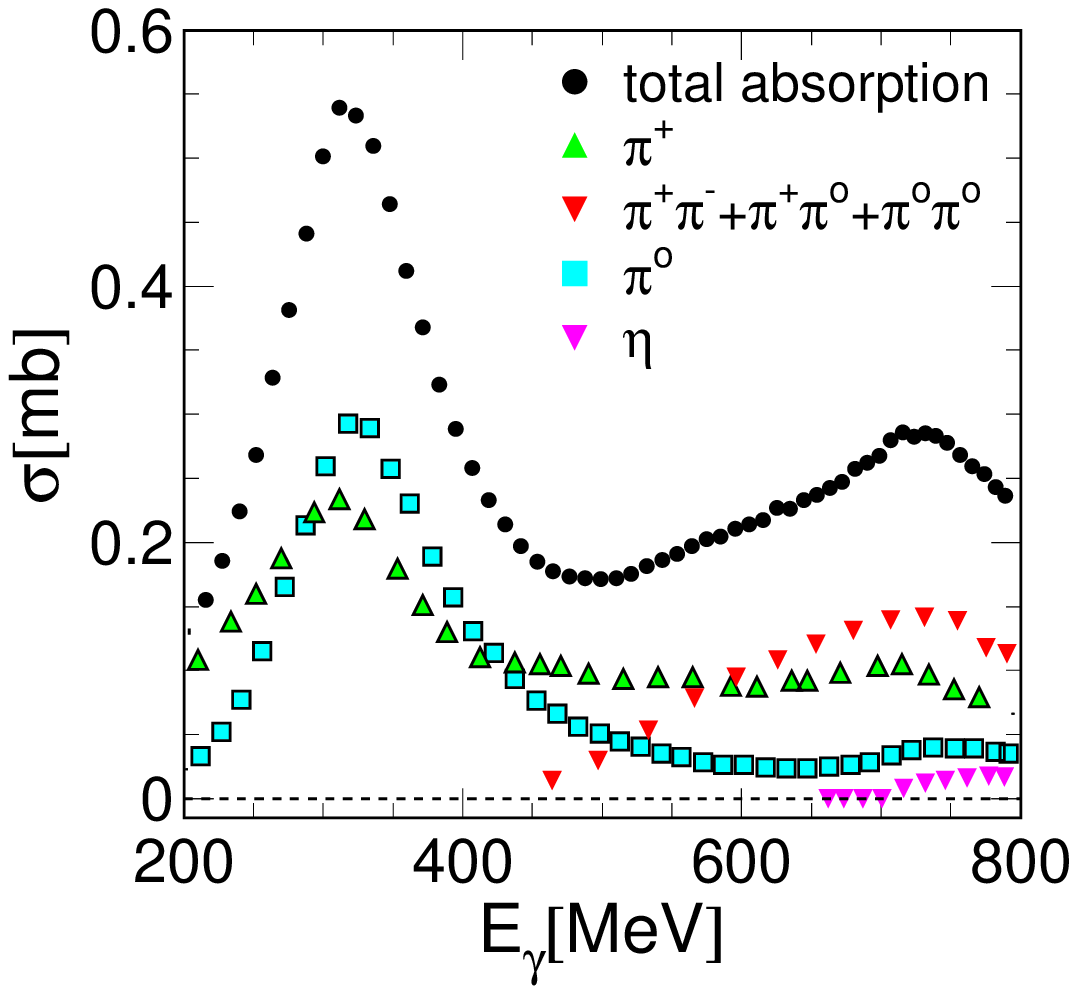}
\end{minipage}
\vspace*{-0.4cm}
\caption{Left hand side: TPA from the proton, the deuteron 
and the average for heavy nuclei \cite{Bianchi_93}. Right hand side: Partial 
cross sections for the proton 
\cite{Krusche_99,Buechler_94,Braghieri_95,Krusche_95,MacCormick_96,Haerter_97,Wolf_00}.}
\label{fig:5}
\end{figure}

Among the clearest experimental observations of in-medium effects is the 
suppression of the second resonance peak in total photoabsorption (TPA) 
\cite{Frommhold_92,Bianchi_93}. TPA on the free proton 
shows a peak-like structure at incident photon energies between 600 and 800 MeV, 
which is attributed to the excitation of the P$_{11}$(1440), D$_{13}$(1520), 
and S$_{11}$(1535) resonances. This structure is not visible for nuclei over 
a wide range of mass numbers from lithium to uranium 
(see fig. \ref{fig:5}, left hand side). 

\vspace*{-0.4cm}
The resonance bump on the free proton consists of a superposition of reaction 
channels with different energy dependencies (see fig. \ref{fig:5}, right hand
side) which complicates the situation \cite{Krusche_03}. Much of the rise of 
the cross section towards the maximum around 750 MeV is due to the double pion
decay channels, in particular to the  n$\pi^o\pi^+$ and p$\pi^+\pi^-$ final
states. A study of the partial reaction channels is thus desirable. 
However, the experimental identification of exclusive final 
states is more involved and FSI effects must be accounted 
for. The interpretation of exclusive measurements therefore always needs 
models which account for the trivial in-medium and FSI effects like absorption 
of mesons and propagation of mesons and resonances through nuclear matter. 

The results for meson photoproduction off the free proton suggest, that
pion and $\eta$ photoproduction are best suited for a comparison of the 
in-medium properties of the D$_{13}$ and S$_{11}$ resonances \cite{Krusche_03}.
The total cross section for $\eta$ photoproduction is completely dominated 
in the second resonance region by the S$_{11}$(1535) resonance
\cite{Krusche_97}. On the other hand, the resonance structure in $\pi^o$ 
photoproduction is strongly dominated by the D$_{13}$(1520) resonance.
Furthermore, the analysis of double pion production from the 
free nucleon has shown that a significant contribution to the decay strength
of the D$_{13}$ resonance \cite{Zabrodin_99,Langgaertner_01} comes from the 
D$_{13}\rightarrow N\rho$ decay. The large broadening of the D$_{13}$
in-matter spectral function predicted in \cite{Post_04} is related to this
channel. In double pion production reactions the $\rho$ contributes to the 
$\pi^o\pi^{\pm}$ and $\pi^+\pi^-$ final states, 
but not to $\pi^o\pi^o$ since $\rho^o\rightarrow\pi^o\pi^o$ is forbidden.  

\begin{figure}[th]
\begin{minipage}{11cm}
\includegraphics[width=0.35\textwidth]{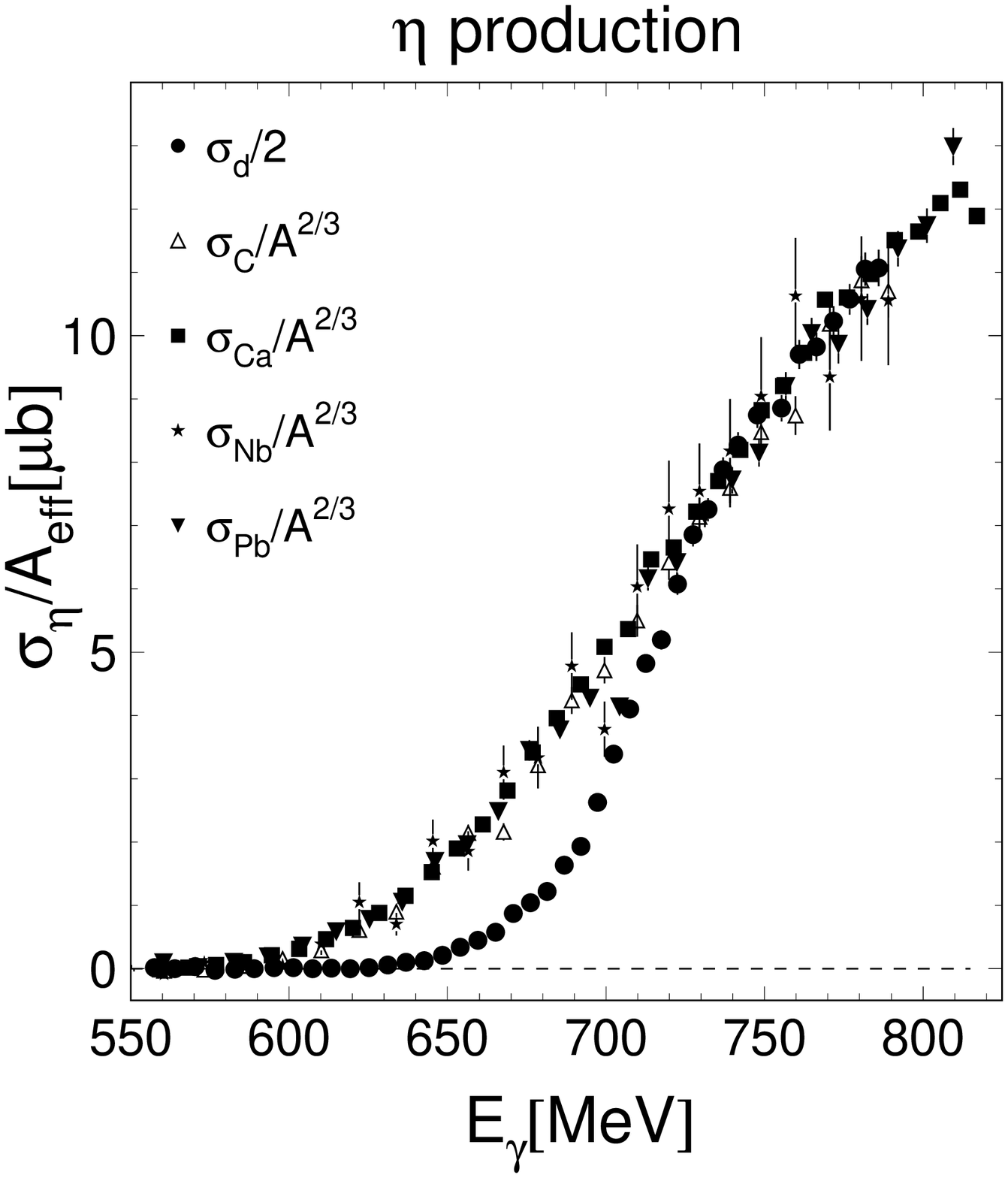}
\includegraphics[width=0.38\textwidth]{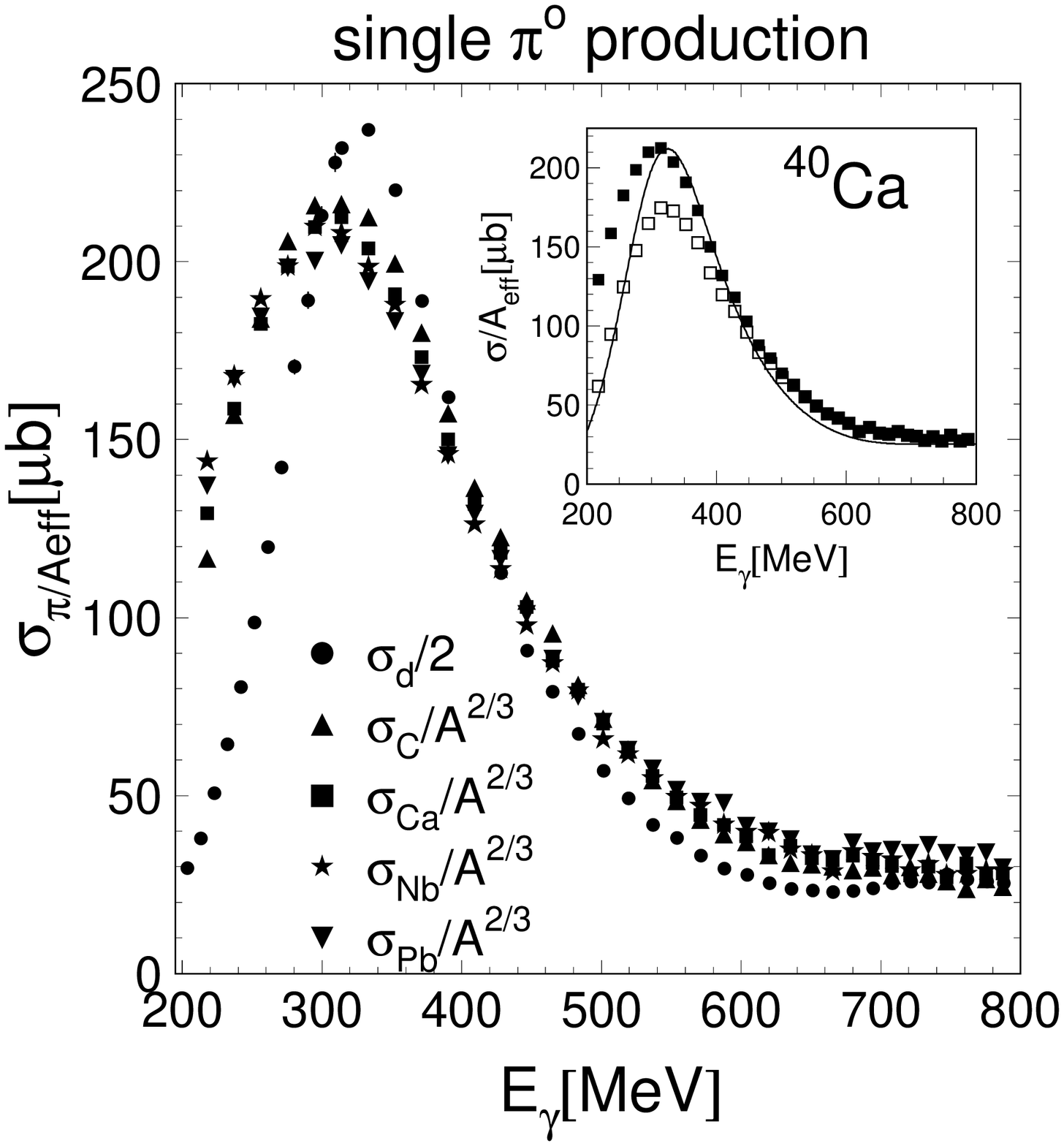}\\
\includegraphics[width=0.37\textwidth]{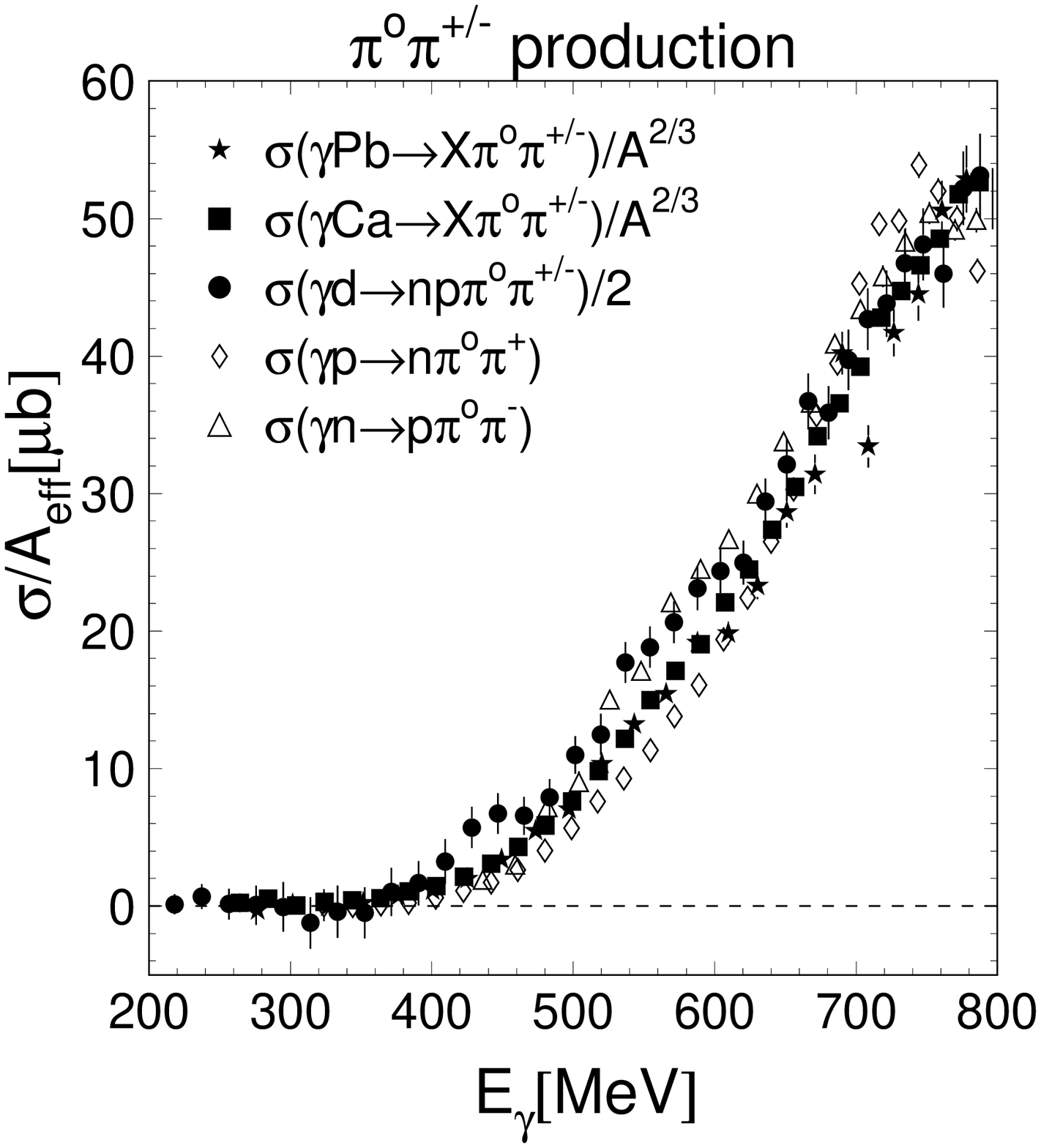}
\includegraphics[width=0.35\textwidth]{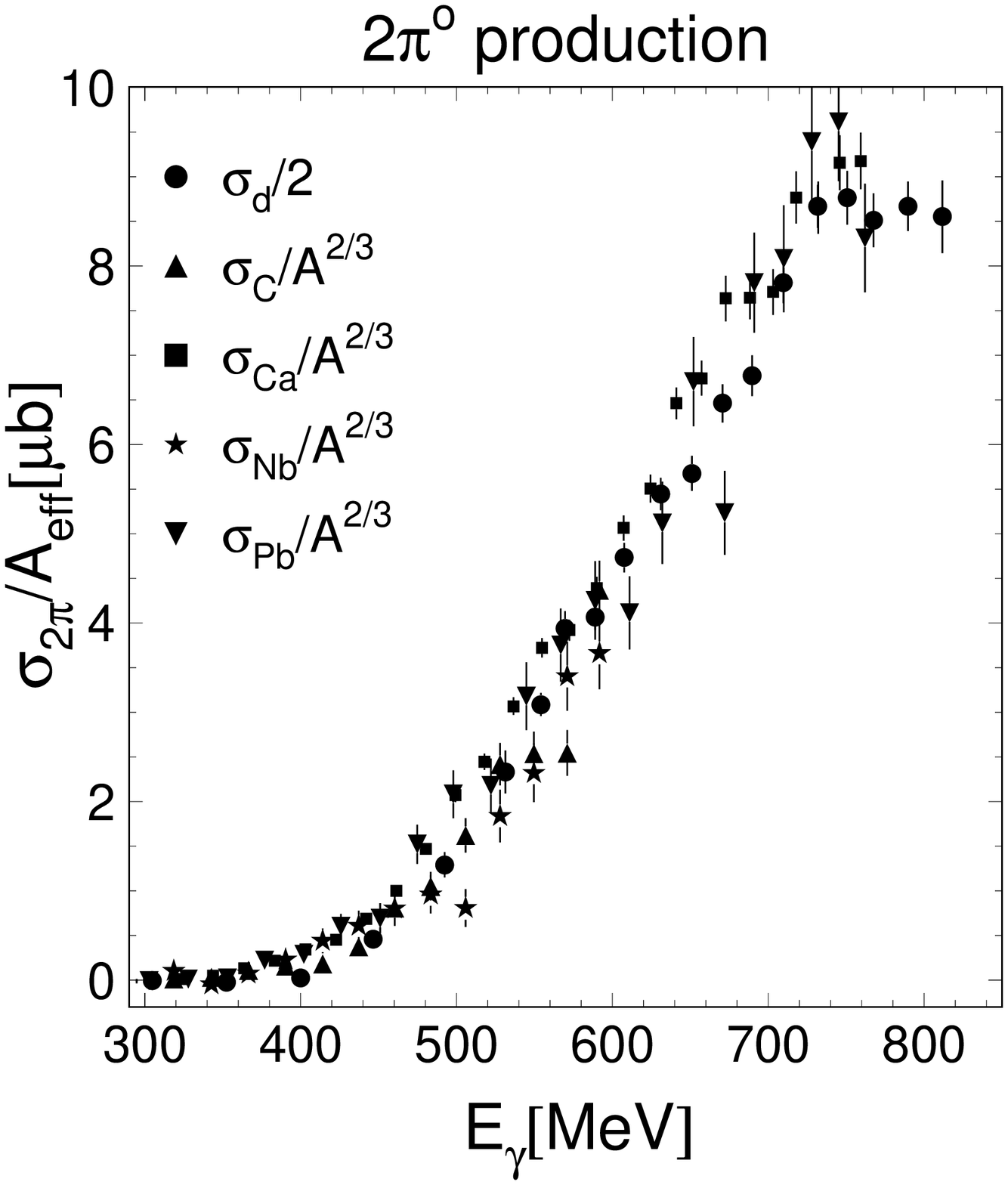}
\vspace*{-9.5cm}
\end{minipage}
\hspace*{8.2cm}
\begin{minipage}{5.cm}
\caption{Total cross sections of quasifree meson photoproduction from nuclei
compared to the respective cross sections from the deuteron. All data are
normalized to $A_{eff}$, where $A_{eff}=A$ for the nucleon and the deuteron
and $A_{eff}=A^{2/3}$ for heavier nuclei. The insert for single pion production
compares the $^{40}$Ca data with (filled symbols) and without (open symbols)
the coherent component to the deuteron response folded with the nucleon momentum
distribution of Ca.
}
\label{fig:6}
\end{minipage}
\end{figure}

\vspace*{0.6cm}
The measured excitation functions for quasifree single $\pi^o$, $\eta$,
$\pi^o\pi^o$, and $\pi^o\pi^{\pm}$ photoproduction from nuclei
\cite{Roebig_96,Krusche_04} are summarized in fig. \ref{fig:6}.
All nuclear cross sections are related to a good approximation 
to the deuteron cross section by:
\begin{equation}
\frac{\sigma_x^{qf}(A)}{A^{2/3}}\approx\frac{\sigma_x^{qf}(d)}{2}
\end{equation}
The scaling among the heavier nuclei holds even when the comparison to the
deuteron is disturbed, e.g. by the effects of coherent single
$\pi^0$ production at low photon energies and or by effects of Fermi smearing
close to the $\eta$ threshold.
The $A^{2/3}$ scaling is the limiting case of strong FSI effects due to the 
short pion mean free path. This means that the quasifree exclusive
reactions probe only the nuclear surface region. Since the properly scaled
nuclear cross sections agree with the deuteron cross section no significant
in-medium effects are observed for the low density surface zone of the nuclei.
Consequently, as already discussed in \cite{Roebig_96,Krusche_01} no direct 
indications for a broadening of the S$_{11}$ or D$_{13}$ resonance have been
found in the exclusive quasifree reactions. 
\begin{figure}[th]
\centerline{
\epsfysize=4.8cm \epsffile{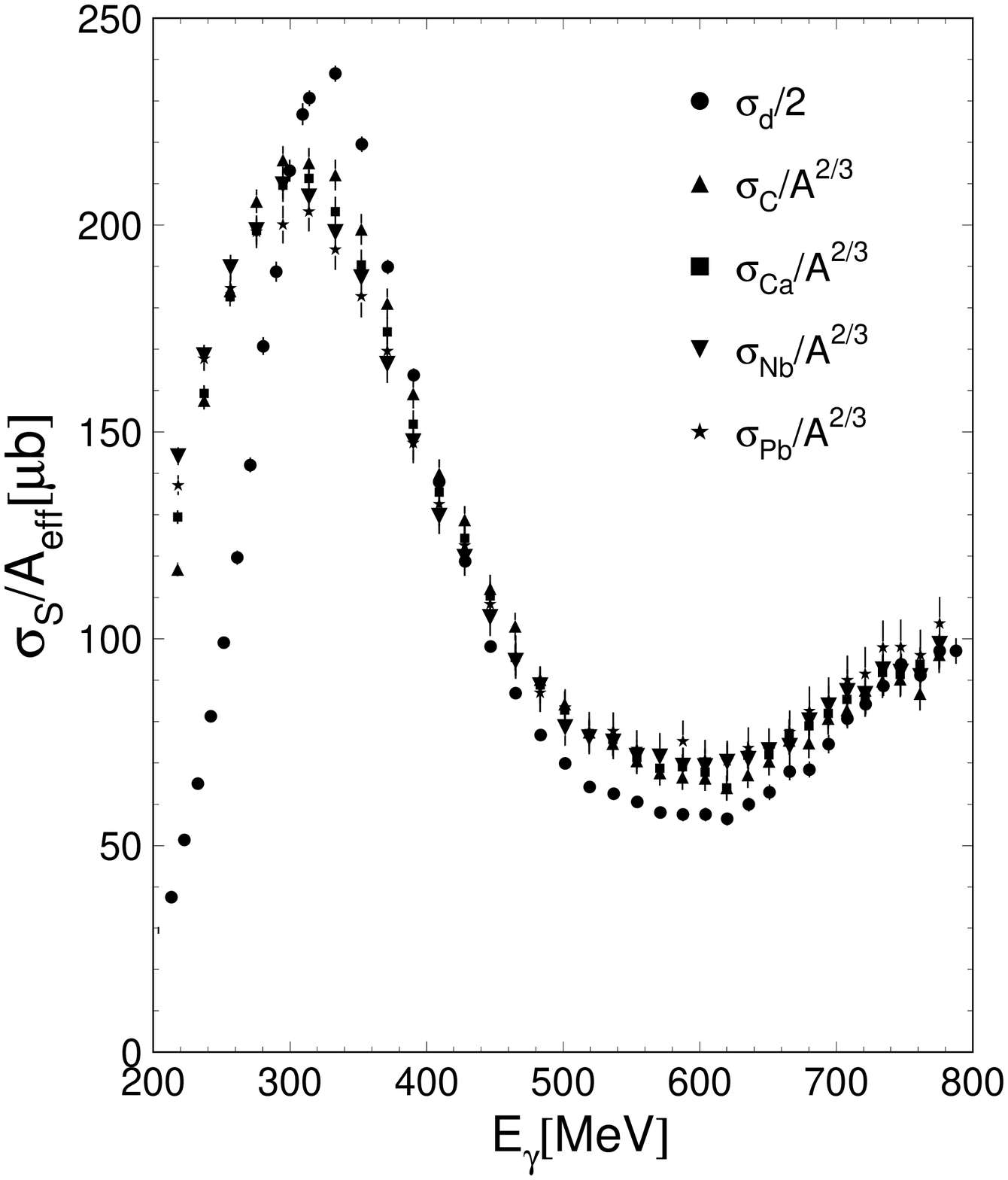}
\epsfysize=4.8cm \epsffile{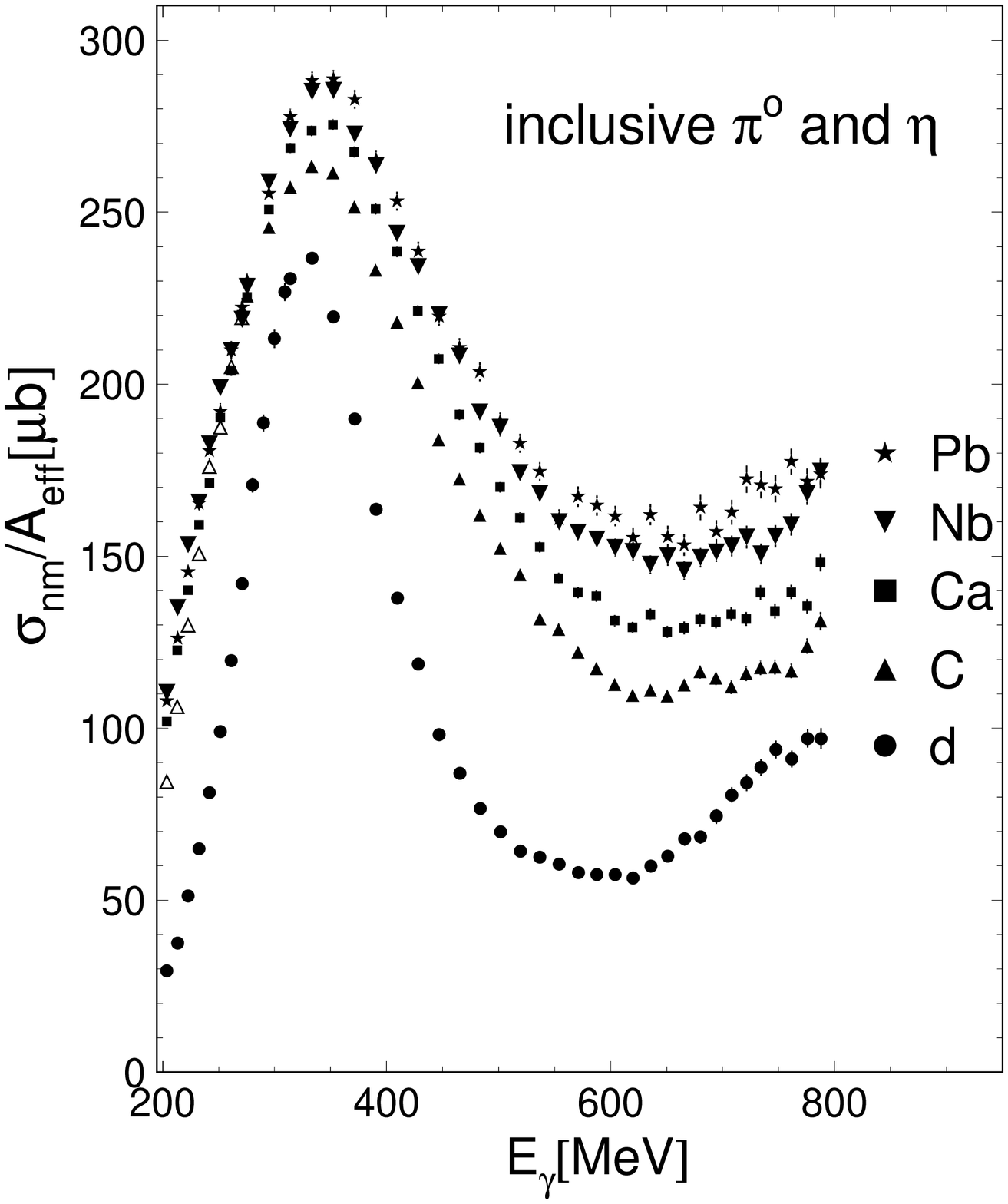}
\epsfysize=4.8cm \epsffile{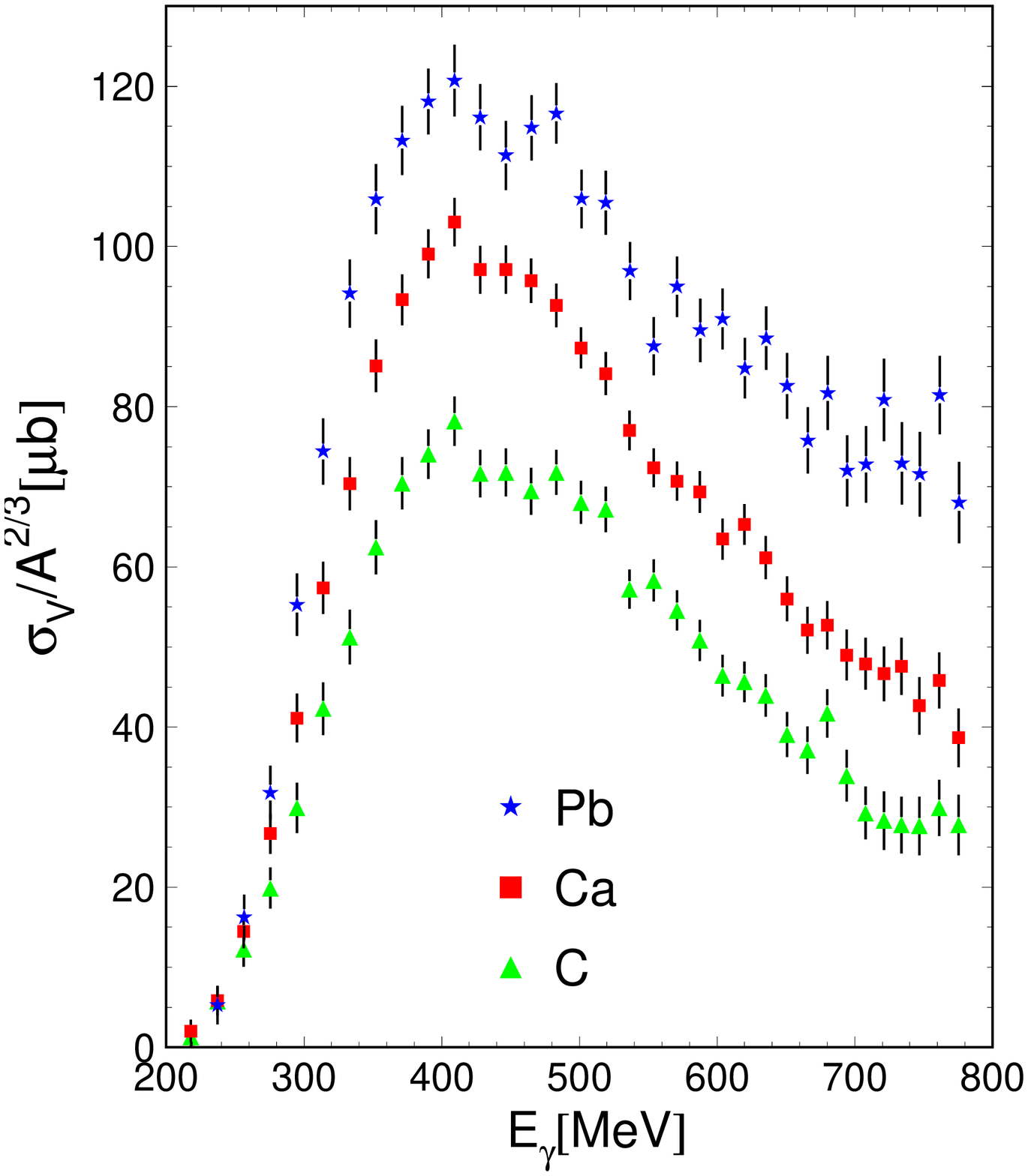}}
\caption{Left hand side: sum of exclusive quasifree and coherent
channels $\sigma_{S}$, middle: total inclusive cross section $\sigma_{nm}$ for 
neutral meson production ($A_{eff}$=2 for the deuteron and $A_{eff}=A^{2/3}$
for $A>2$),  right hand side: non-quasifree components $\sigma_V$.
}
\label{fig:7}
\vspace*{-0.3cm}
\end{figure}
This is particularly obvious for 
the cross section sum $\sigma_{S}$ of these reactions:
\begin{equation}
\sigma_{S} = 
\sigma_{\pi^o}^{qf}+\sigma_{\eta}^{qf}
+\sigma_{2\pi^o}^{qf}+\sigma_{\pi^o\pi^{\pm}}^{qf}
\end{equation}
which is shown in fig. \ref{fig:7}, (left hand side).
The behavior of $\sigma_S$ throughout the second resonance is very similar for
the deuteron and the heavy nuclei. The resonance structure is almost 
identical, no in-medium effects are visible, and the scaling indicates the 
dominance of FSI effects.

The total inclusive cross section of $\pi^o$ and $\eta$ photoproduction
$\sigma_{nm}$ (see fig. \ref{fig:7}, middle) was also 
extracted in \cite{Krusche_04}. It includes reactions where 
a single neutral pion is observed which does not fulfill the kinematic
constraints of quasifree or coherent reactions. These are mainly reactions with
strong FSI, e.g. double pion production with one pion re-absorbed in the
nucleus. The behavior is somewhere in between total photoabsorption and the 
quasifree component. The resonance structure is still visible for heavy nuclei,
but it is much less pronounced than for the deuteron. The difference 
$\sigma_V=\sigma_{nm}-\sigma_S$ between the inclusive cross section 
and the quasifree components (see fig. \ref{fig:7}, right hand side) has
a completely different energy dependence. It does not show any indication 
of the second resonance bump.

The mass number scaling of the different components of the total neutral meson
production cross section was analyzed in \cite{Krusche_04a} with the simple
ansatz $\sigma (A)\propto A^{\alpha}$. In case of the quasifree component 
$\sigma_S$ the exponent $\alpha$ was found to be close to 2/3 over the whole 
energy range. This is the expected behavior of surface dominated meson 
production. However, $\alpha$ is significantly larger for the non-quasifree 
components, which 
indicates that this contribution probes to some extent the nuclear volume. 
In this case, the appearance of the second resonance peak in $\sigma_S$ and 
its complete suppression in $\sigma_V$ could indicate a strong density 
dependence of the effect. 

\section*{Acknowledgments}
The discussed results are part of the experimental program of the TAPS
collaboration. I like to acknowledge in particular the contributions of 
F. Bloch, S. Janssen, and M. R\"obig-Landau. This work was supported by the 
Swiss National Fund.

\vfill\eject
\end{document}